# Making Sense of the Evolution of a Scientific Domain: A Visual Analytic Study of the Sloan Digital Sky Survey Research

CHAOMEI CHEN[a], JIAN ZHANG[a], MICHAEL S. VOGELEY[b]
[a]College of Information Science and Technology, Drexel University, Philadelphia, PA 19104-2875 (USA)
[b]Department of Physics, College of Arts and Sciences, Drexel University, Philadelphia, PA 19104-2875 (USA)
Email: {chaomei.chen, jz85, vogeley}@drexel.edu

**ABSTRACT**
We introduce a new visual analytic approach to the study of scientific discoveries and knowledge diffusion. Our approach enhances contemporary co-citation network analysis by enabling analysts to identify co-citation clusters of cited references intuitively, synthesize thematic contexts in which these clusters are cited, and trace how research focus evolves over time. The new approach integrates and streamlines a few previously isolated techniques such as spectral clustering and feature selection algorithms. The integrative procedure is expected to empower and strengthen analytical and sense making capabilities of scientists, learners, and researchers to understand the dynamics of the evolution of scientific domains in a wide range of scientific fields, science studies, and science policy evaluation and planning. We demonstrate the potential of our approach through a visual analysis of the evolution of astronomical research associated with the Sloan Digital Sky Survey (SDSS) using bibliographic data between 1994 and 2008. In addition, we also demonstrate that the approach can be consistently applied to a set of heterogeneous data sources such as e-prints on arXiv, publications on ADS, and NSF awards related to the same topic of SDSS.

## Introduction

Analyzing the evolution of a scientific field is a challenging task. Analysts often need to deal with the overwhelming complexity of a field of study and work back and forth between various levels of granularity. Although more and more tools become available, sense making remains to be one of the major bottleneck analytical tasks. In this article, we introduce a new visual analytic approach in order to strengthen and enhance the capabilities of analysts to achieve their analytical tasks. In particular, we will focus on the analysis of co-citation networks of a scientific field, although the procedure can be applied to a wider range of networks.

## Analyzing Dynamic Networks

Many phenomena can be represented in the form of networks, for example, friendship on FaceBook, trading between countries, and collaboration in scientific publications (Barabási, et al., 2002; Snijders, 2001; Wasserman & Faust, 1994). A typical path of analyzing a dynamic network may involve the following steps: formulate, visualize, clustering, interpret, and synthesize (See Figure 1). Many tools are available to support these individual steps. On the other hand, analysts often have to improvise different tools to accomplish their tasks. For example, analysts may divide the nodes of a network into clusters by applying a clustering algorithm to various node attributes. Clusters obtained in such ways may not match the topological structure of the original network, although one may turn such discrepancies into some good use. We are interested in processes that would produce an intuitive and cohesive clustering given the topology of the original network.
The new procedure we are proposing is depicted in Figure 1b. It streamlines the key steps found in a typical path. The significance of the streamlined process is that it determines clusters based on the strengths of the links in the network. In Figure 1c, we show that the new procedure leads to several advantages such as increased clarity of network visualization, intuitive aggregation of cocited references, and automatically labeling clusters to characterize the nature of their impacts.

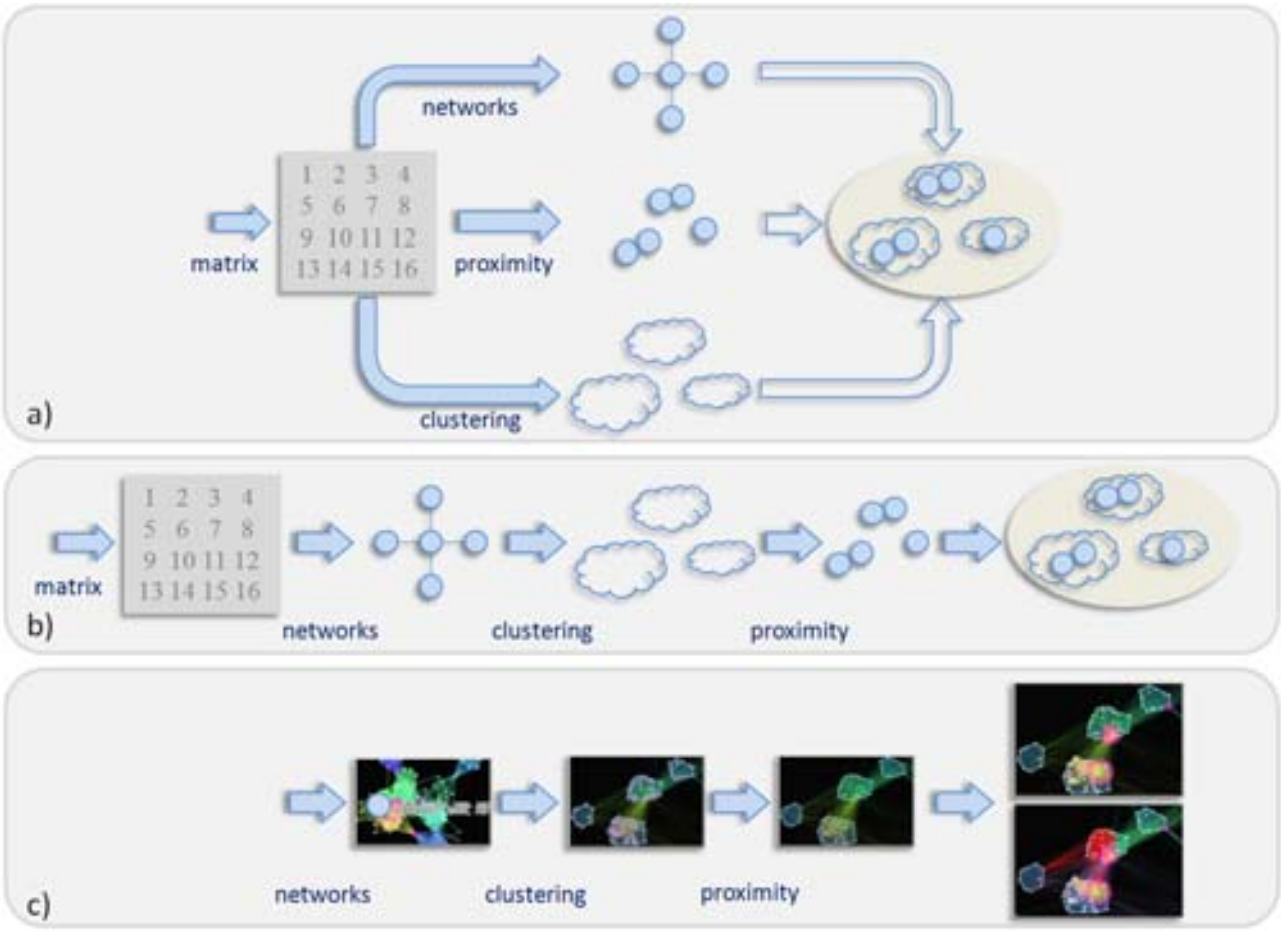

**Figure 1. A common path of network analysis (a) and a new procedure (b) and its effects (c).**

According to Gestalt principles, perceived proximity plays a fundamental role in how we aggregate objects into groups (Koffka, 1955). If some objects appear to be closer to each other than the rest of objects, we tend to be convinced that they belong together. Seeing objects in groups instead of individual objects is important in many cognitive and analytical activities. As a generic chunking method, we often use it to simplify a complex phenomenon so that we can begin to address generic properties.

Figure 2 shows three illustrative examples of how clarity of displayed proximity can make the chunking task easy (Figure 2a) or hard (Figure 2c). Co-citation networks represent how often two bibliographic items are cited together, for example, authors in author co-citation networks (White & Griffith, 1981) and papers in document co-citation networks (Small, 1973). When analyzing co-citation networks, or more generic networks, we often find ourselves in the situation depicted in Figure 2c. Our goal is to find mechanisms that can improve the representation and approach the ideal case of Figure 2a. A hot topic in the graph drawing community, called constraint graph drawing, addresses this problem (Dwyer, et al., 2008). In this article, however, we propose an alternative solution that is in harmony with our overall goal for strengthening visual analytical capabilities of analysts.

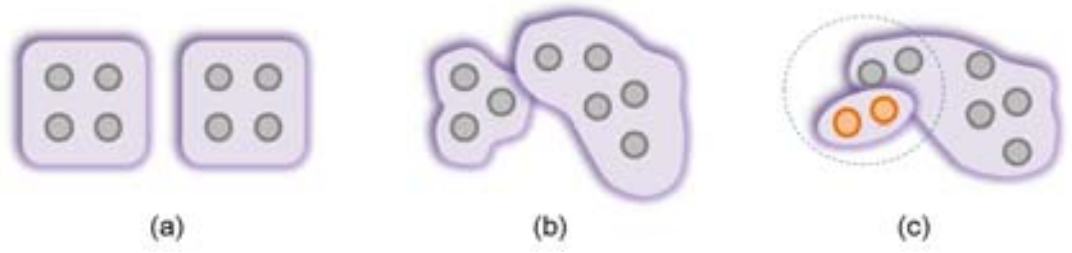

**Figure 2. Clarity of displayed proximity plays an important role in chunking tasks.**

## Methods

The proposed procedure consists of the following key steps: constructing an integrated network of multiple networks, finding clusters of nodes in the network based on connectivity, selecting candidate labeling terms for each cluster, and visual exploration and analysis. In this article, we will focus on the new steps, namely clustering, labeling, and visual analysis.

### *Networks*

It is possible to construct a wide range of networks from bibliographic data. For example, CiteSpace supports collaboration networks of coauthors, collaboration networks of institutions and countries, author co-citation networks, document co-citation networks, concept networks of noun phrases and keywords, and hybrid networks that consist of multiple types of nodes and links (Chen, 2006). For simplicity, we will primarily focus on document co-citation networks and relevant analysis.

The study of an evolving scientific field often needs to focus on how the field evolves over several years. The notion of progressive knowledge domain visualization was introduced in order to accommodate such needs (Chen, 2004). Suppose we are interested in the evolution of a field in a time interval T, for example, between 1990 and 2008, time slicing is an operation that divides the interval T into multiple equal length sub-intervals $T_i$. While CiteSpace implements non-overlapping sub-intervals, overlapping sub-intervals may be also considered. For each sub-interval, or time slice, a network can be constructed purely based on data falling into the time slice. For example, a co-citation network of 1990 can be constructed based on instances of co-citation found in scientific papers published in 1990. Similarly, a collaboration network of authors of 1990 would consist of researchers who have published together in 1990.

In this article, the goal of our visual analysis is to assess the impact of the Sloan Digital Sky Survey (SDSS) (Chen, Zhang, Zhu, & Vogeley, 2007). The input data was retrieved from the Web of Science with a topic search of articles on SDSS between 1994 and 2008. We used one year as the length of our time slice. In each time slice, a co-citation network was constructed based on the co-citation instances made by the top 30 most cited records published in the corresponding time interval. The top 30 most cited records were determined by their original times cited in the Web of Science. These individual networks led to an integrated network of 259 nodes and 2,130 co-citation links. Our subsequent study will focus on this network.

### *Clustering*

We utilize the spectral clustering technique to identify clusters in networks. Spectral clustering has many fundamental advantages over the traditional clustering algorithms such as *k*-means or single linkage. For example, results obtained by spectral clustering very often outperform the traditional approaches (Luxburg, 2006; Ng, Jordan, & Weiss, 2002; Shi & Malik, 2000).

There are many reasons one might need to identify clusters in data given in the form of associative networks, for example, to find communities in a social network (Girvan & Newman, 2002). In such situations, the problem of clustering can be stated as the need to finding a partition of the network such that nodes within a cluster would be tightly connected, whereas nodes between different clusters would be loosely connected or not connected at all. Consider our document co-citation network, this is equivalent to find a partition such that references within a cluster would be significantly more cocited than references from different clusters. Spectral clustering offers a solution to such graph partitioning problems. This view of clustering fits our needs perfectly and intuitively. In addition, since spectral clustering comes naturally for a network, it has the distinct advantage over alternative clustering algorithms that rely on node attributes rather than linkage. Compared to traditional linkage-based algorithms such as single linkage, spectral clustering has the advantage due to its linear algebra basis. Spectral clustering is implemented as an approximation to the graph partitioning problem with constraints stated above, i.e. members within clusters are tightly coupled, whereas members between clusters are loosely connected or disconnected.

### *Enhancing the Clarify of Layout*

As a welcome by-product of spectral clustering, we enhance the clarity of network visualization by taking into account the graph partitioning information. Constrained graph drawing is currently a hot topic. The goal is to layout a graph with given constraints (Dwyer, et al., 2008). Given a graph partition, drawing the graph with minimal overlapping partition regions is one of the common special cases.

One of the common analytical tasks in network analysis is to study the largest connected component of a network. The ability to find finer-grained clusters has significant theoretical and practical implications. Our previous studies show that co-citation networks may contain tightly knitted components. In other words, if the largest connected component is densely connected, it would be hard to identify meaningful sub-structures. Since spectral clustering works at the strength of links rather than the simple presence or absence of links, we expect that spectral clustering will find finer-grained clusters even within large connected components.

We make simple modifications of force-directed graph layout algorithms to improve the clarity of such processes. A major advantage of using spectral clustering is that it can be applied to any network because structural information is all it needs. Briefly speaking, once the clustering information is available, the layout algorithm would maintain the strength of a within-cluster link but downplay or simply ignore a between-cluster link during the layout process.

## Cluster Labeling

Once clusters are identified in a network, the next step is to help analysts make sense of the nature of these clusters, how they connect to one another, and how their relationships evolve over time. We introduce algorithmic cluster labeling to assist this step.

### *Methodological Issues*

Traditionally, clusters would be identified using an independent clustering process in contrast to the integrative and cohesive approach we described above. Traditionally, sense making identified clusters is essentially a manual process. Researchers often examine members of each cluster and sum up what they believe to be the most common characteristics of the cluster. There are two potential drawbacks with the traditional approach, especially in the study of co-citation networks. First, co-citation clusters could be too complex to lend themselves to simple eyeball examinations. The cognitive load required for aggregating and synthesizing the details is likely to be high. A computer-generated baseline list of candidate labeling terms would reduce the burden significantly. Second, and more importantly, studying co-citation clusters themselves does not necessarily reveal the actual impacts of these clusters. In fact, it is quite possible that co-citation clusters are referenced by subsequent publications not only in the same topical area, but also in topical areas that may be not obvious from the cited references alone. In other words, traditional studies often infer the nature of co-citation grouping, but they do not directly address the question of why a co-citation cluster is formed in the first place.

Unlike traditional studies of co-citation networks, we focus on the citers to a co-citation cluster instead of the citees and label the cluster according to salient features selected from the titles and index terms of the citers. Our prototype implements a number of classic feature selection algorithms, namely term frequency by inverse document frequency (tf*idf) (Salton, Allan, & Buckley, 1994), log-likelihood ratio test (Dunning, 1993), mutual information (not discussed in this article), and latent semantic indexing (Deerwester, Dumais, Landauer, Furnas, & Harshman, 1990). Formal evaluations are beyond the scope of this article. As part of future work, we are planning cross-validations with labels generated by other means and a study of topological distributions of labels selected by different algorithms in networks of terms.

### *Selection by tf*idf*

Given a cluster C, the citing set consists of articles that cite one or more members of the cluster. Candidate labeling terms for the clusters are selected from the titles, abstracts, or index terms of articles in the citing set. In this article, we focus on selecting labels from titles and index terms. First, we extract noun phrases from titles and compute weights of these phrases using tf*idf. A noun phrase may consist of a noun and possibly modified by one or more adjectives, for example, supermassive black hole. Plurals are stemmed using a few simple stemming rules. Using tf*idf has known drawbacks due to the term independency assumption. Nevertheless, its properties are widely known; this, it serves as a good reference point.

*Selection by log-likelihood ratio test*

The log-likelihood ratio test we adapted in our approach measures how often a term is expected to be found within a cluster's citer set to how often it is found within other clusters' citer sets. It tends to identify the uniqueness of a term to a cluster.

*Selection by latent semantic indexing*

Latent semantic indexing, or latent semantic analysis (LSA), is another classic method for dimension reduction in text analysis. LSA utilizes the singular value decomposition (SVD) technique on a term by document matrix. In order to select candidate labeling terms of a cluster, we select the top 5 terms with the strongest coefficients on each of the first and second dimensions of the latent semantic space derived from the citer set of the cluster.

*Expanding the scope of the analysis to multiple sources*

We expand the scope of our understanding of the subject domain by applying the same method to multiple sources that usually do not have citation information in readily formatted forms but nevertheless are important in the development of a scientific domain. In particular, the following additional sources are studied:

- *arXiv:* e-prints posted to the arXiv's astro-ph during the last 15 days of the time of writing (August 22, 2009). Note that this is not equivalent to all the SDSS-related e-prints.
- *ADS*: SDSS related papers appeared during January-August 22, 2009
- *NSF Awards*: retrieved with a simple search for SDSS in the descriptions of awards

CiteSpace provides adaptors to process data retrieved from these sources. Since these datasets have no citation information, we first extract noun phrases from these records, then generate clusters and automatically assign labels to clusters. Unless explicitly stated otherwise, cluster labels are selected by log-likelihood ratio from title terms. We expect this approach will be valuable for analysts to compare emerging patterns from related but distinct data sources.

## Results

This section has two parts: the results of cocitation networks and the results of networks of co-occurring terms from multiple data sources such as arXiv eprints and NSF awards.

*Clusters in Cocitation Networks*

First, we show how spectral clustering can enhance the clarity of network visualization. In Figure 3, the left image shows a visualization of the core of our SDSS co-citation network, the right image shows a cluster-enhanced visualization of the same network. Before the enhancement, York-2000 and Fukugita-1996 appear to be very close to each other. After the enhancement, it becomes clear that they belong to two distinct co-citation clusters. The improved clarity will be useful in the subsequent analysis of the domain.

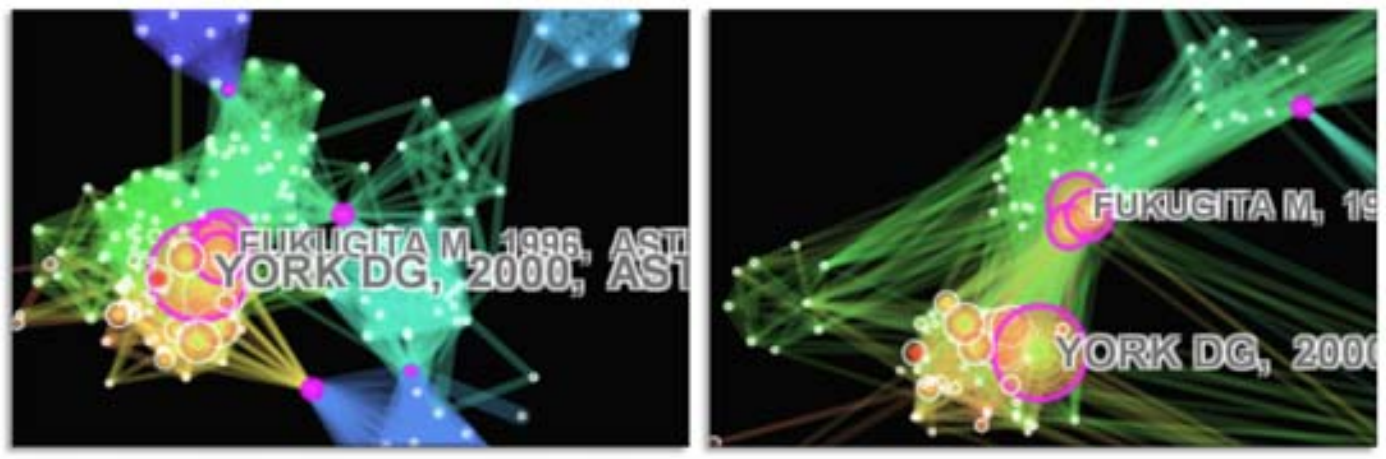

**Figure 3. The effect of enhanced clarity (left: before; right: after).**

The four images in Figure 4 show various options of visual exploration. The two clusters in the middle now become separated from each other. Despite the numerous links between the two clusters, spectral clustering detected that they are two distinct clusters in terms of how they are cocited. The two images on the second row depict pivotal nodes (with purple rings) and nodes with citation burst (with red tree rings). The pivotal nodes play a brokerage role between different clusters. They are particularly useful in interpreting the macroscopic structure of a knowledge domain (Chen, 2006). The red lines in the lower right image depict co-citation instances made in a particular time slice, in this case, year 2001. These red lines show that the middle cluster is essentially formed in 2001 with between-cluster co-citation links to two neighboring clusters. These features would allow analysts to pin point the specific time when attention is paid to a cluster and how multiple clusters are connected.

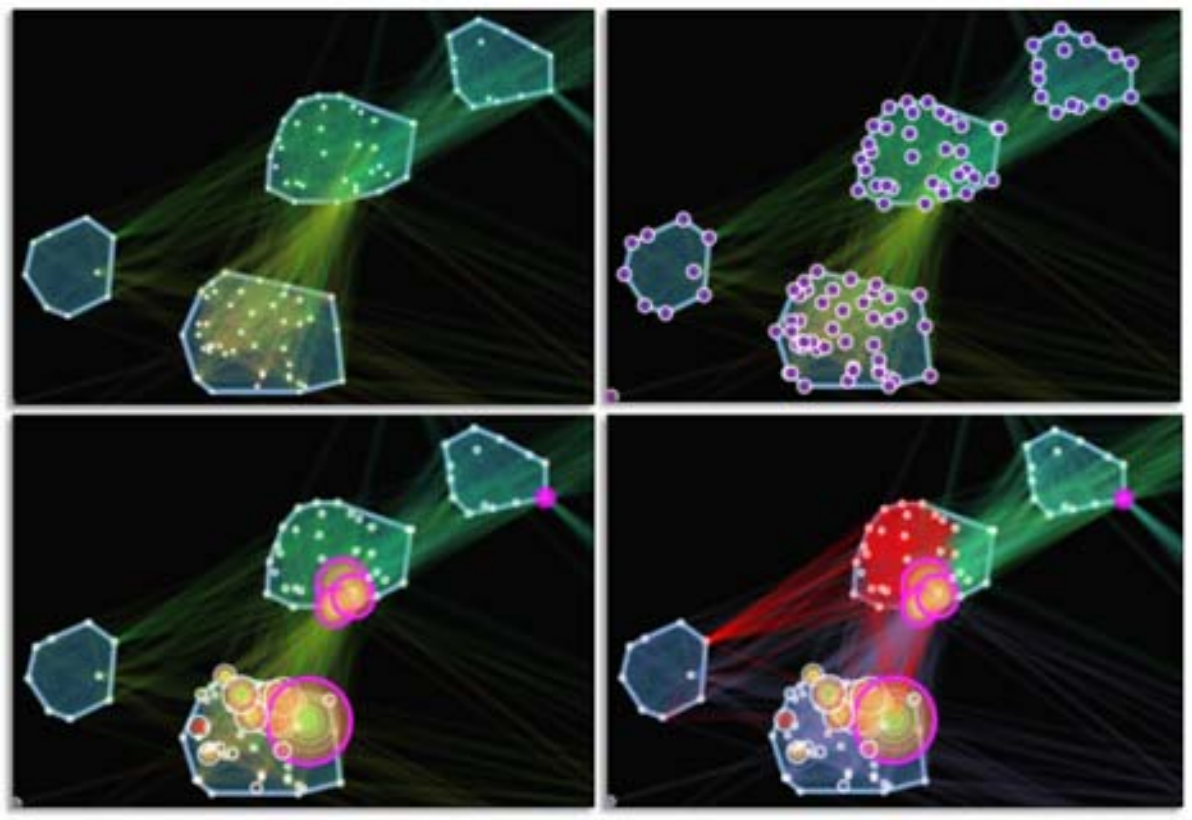

**Figure 4. Various node and link attributes depicted for visual exploration.**

Figures 5 and 6 are screenshots of the region of some of the core clusters of the SDSS co-citation network. In Figure 5, clusters are labeled by title phrases selected by tf*idf. Four clusters (#9, #10, #11, and #12) have the identical label of “sloan digital sky survey.” The numbers in front of the labels are weights of the corresponding labels by tf*idf. The clarity of the layout is enhanced by spectral clustering. On the one hand, it appears that the four clusters are common enough to get the same label. On the other hand, we also know that each of them must play a unique role in the subsequent course of the field because they are separated by how they are cocited by researchers of the SDSS field.

Figure 6 shows the same clusters but with labels selected by a different algorithm, i.e. the log-likelihood ratio test. The four clusters now have different labels. Note other clusters’ labels are changed too. Cluster 9 is labeled as field methane dwarf. Methane dwarfs are very cool brown dwarfs. They are smaller than a star, but larger than a planet, and they are very hard to detect because they are very faint in the sky[1]. Finding rare objects such as methane dwarfs is one of the first discoveries made possible by the SDSS survey. Cluster 10 is labeled as high-redshift quasar. The redshift measures how far the light of an astronomical object has been shifted to longer wavelengths due to the expansion of the Universe. The higher the redshift, the more distant the astronomical object. Finding high-redshift quasars is important for the study of the early evolution of the Universe. Cluster 11 is labeled as dust emission. Our subsequent analysis shows that the broader context of this cluster is dust emission from quasars. Cluster 12 is labeled as luminous red galaxy. This cluster is in fact the largest cluster in the SDSS co-citation network, concerning various properties of galaxies.

[1] http://www.sdss.org/news/releases/19990531.dwarf.html

In summary, labels selected by log-likelihood ratio test appear to characterize the nature of clusters with finer-grained concepts than labels selected by tf*idf. Specific labels are useful for differentiating different clusters, whereas more generic labels tend to be easy to understand, especially for domain novices.

The third labeling algorithm is based on latent semantic analysis (LSA). Unlike the first two labeling algorithms, the LSA-based labeling algorithm uses single words instead of multi-word noun phrases. The LSA-based labeling algorithm first identifies the primary and secondary dimensions of the latent semantic space derived from the citer set of each cluster. Next, it selects the top 5 terms with the strongest weights along each dimension. Table 1 lists the selected terms for the four clusters discussed above. The primary concept terms appear to correspond to the noun phrase labels identified by tf*idf. The secondary concept terms appear to be more specific. Taken these terms together for each cluster, we can tell that Cluster 9 is about methane dwarfs, Cluster 10 about quasars, Cluster 11 also about quasars, and Cluster 12 about galaxies. The largest 10 clusters are summarized in Table 2.

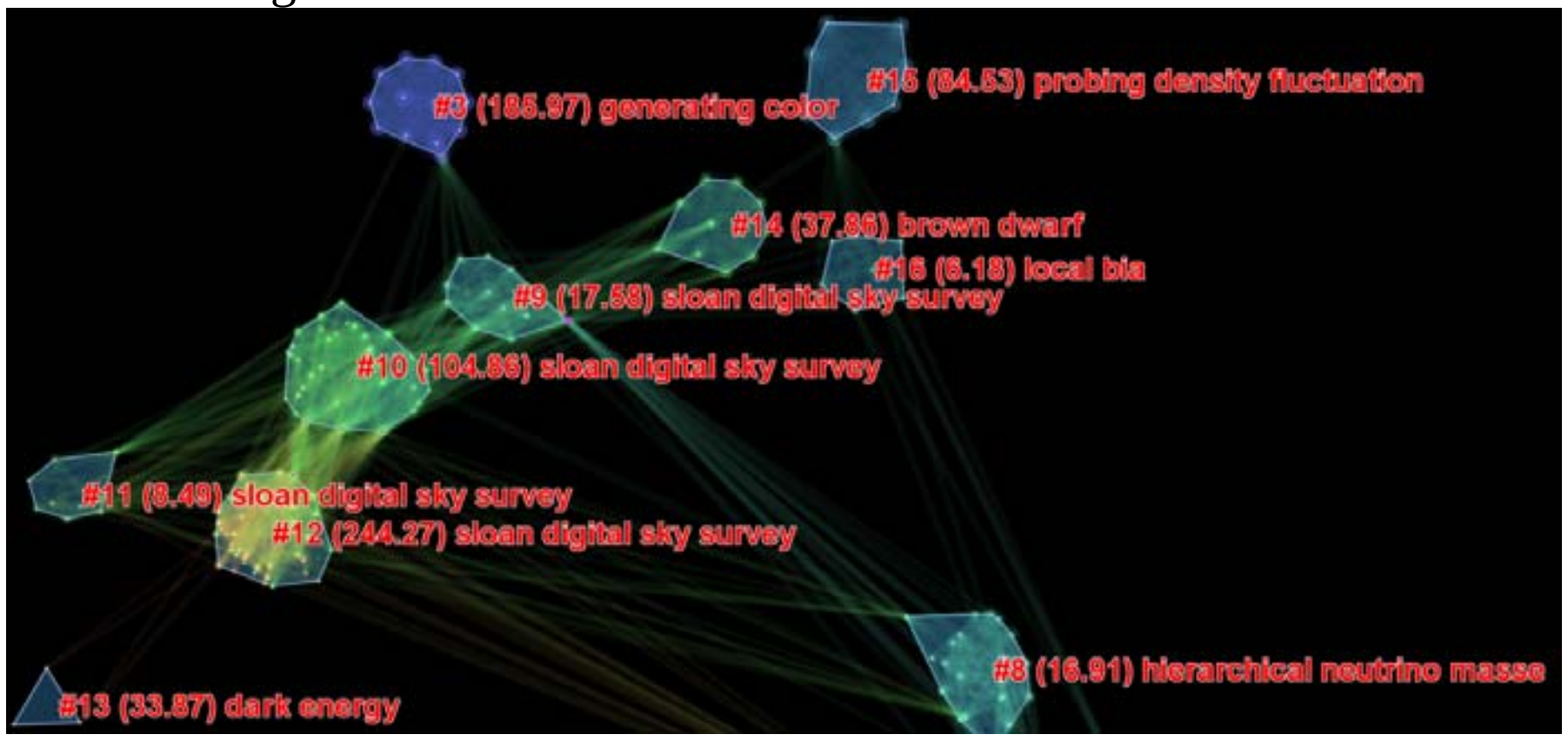

**Figure 5. SDSS-core clusters (#9, #10, #11, #12) are separated but still labeled by tf*idf with the same label *sloan digital sky survey*.**

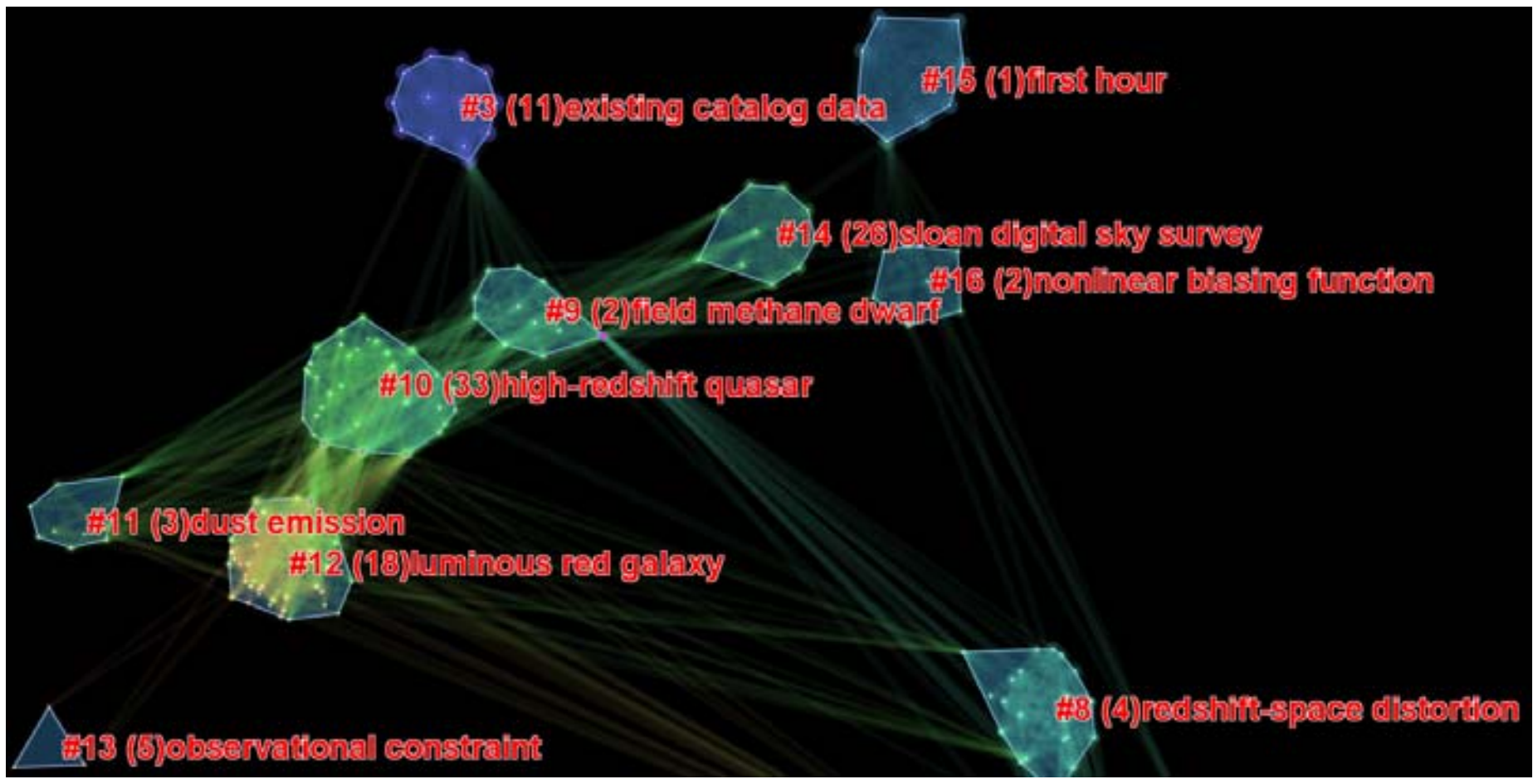

**Figure 6. The four SDSS-core clusters (#9, #10, #11, #12) now have finer-grained labels by log-likelihood ratio tests.**

In the rest of the article, we will triangulate labels selected by the three algorithms and examine the full titles of the most representative citing papers to determine the most appropriate labels of clusters.

**Table 1. Labels by LSA-based selection.**

| Clusters | #9 | #10 | #11 | #12 |
|---|---|---|---|---|
| Primary concept | 4.58 survey | 7.25 survey | 3.60 quasars | 12.52 survey |
| | 4.40 sky | 7.08 sky | 3.43 survey | 12.46 sky |
| | 3.81 sloan | 6.59 sloan | 3.31 sloan | 12.15 sloan |
| | 3.81 digital | 6.59 digital | 3.31 digital | 12.06 digital |
| | 1.87 commissioning | 3.24 quasars | 3.31 sky | 4.44 galaxies |
| Secondary concept | 1.98 methane | 0.72 data | 1.25 quasar | 1.99 sky |
| | 1.40 dwarfs | 0.70 sloan | 1.09 data | 1.78 digital |
| | 1.29 field | 0.70 digital | 0.84 sloan | 1.75 sloan |
| | 1.19 discovery | 0.55 sky | 0.84 digital | 0.80 survey |
| | 1.11 dwarf | 0.47 stars | 0.84 sky | 0.33 quasars |

A total of 22 co-citation clusters were found by spectral clustering. Table 2 shows the 10 largest clusters in terms of the number of references N. The first column (#) shows the cluster IDs. We applied the three labeling methods to the titles and index terms of citing papers of each cluster. The last column shows labels we chose subjectively based on the information shown in all other columns. The numbers in tf*idf columns are the term weights, e.g. (60.78) brown dwarf. The numbers in log-likelihood columns are the frequency of the corresponding terms. For example, (18) luminous red galaxy means that the term appears 18 times in the citer set. The numbers in the Most representative citers column are the number of cluster members the paper cites. For example, the (13) in front of the first title in Cluster 12 means that the paper cites 13 of the 45 members of the cluster. The table is sorted by cluster size.

**Table 2. The largest 10 clusters in the SDSS co-citation network.**

| # | N | Title Terms | | Index Terms | | Titles | Title Terms | Overall |
|---|---|---|---|---|---|---|---|---|
| | | tf*idf | log-likelihood | tf*idf | log-likelihood | Most representative citers | LSA | subjective |
| 12 | 45 | (244.27) sloan digital sky survey (155.49) sky survey (117.05) data release (103.61) active galactic nuclei | (18) luminous red galaxy | (60.78) brown dwarf (48.56) gliese 229b | (38) gliese 229b | (13) the broadband optical properties of galaxies with redshifts 0.02 z 0.22 | 12.52 survey 12.46 sky 12.15 sloan 12.06 digital 4.44 galaxies | Properties of galaxies in SDSS |
| 10 | 33 | (104.86) sloan digital sky survey (77.04) commissioning data (71.24) sky survey (56.26) high-redshift quasar | (33) high-redshift quasar | (819.91) release (809.61) data release | (232) survey photometric system | (15) discovery of a pair of z=4.25 quasars from the sloan digital sky survey | 7.25 survey 7.08 sky 6.59 sloan 6.59 digital 3.24 quasars | SDSS discoveries: high-redshift quasar |
| 8 | 23 | (16.91) hierarchical neutrino masse (12.36) cosmic statistic (11.95) redshift-space correlation function | (4) redshift-space distortion | (80.45) mission; (69.36) quasar | (68) luminosity function | (6) the gravitational lensing in redshift-space correlation functions of galaxies and quasars | 2.65 sky 2.40 survey 2.15 sloan 2.15 digital 1.09 microwave | Redshift-space correlation functions |
| 7 | 22 | (10.3) sloan digital sky survey | (2) background qso | (26.67) cosmology (25.48) | (16) emission | (4) a merged catalog of clusters of galaxies from early sloan digital sky survey data | 3.65 survey 3.08 sky 3.08 sloan | Galaxy clusters in SDSS |
| | | (9.59)<br>early sloan digital sky survey data<br>(9.59)<br>merged catalog<br>(7.97)<br>galaxy cluster | | stars | | (4) clustering and large-scale structure with the sloan digital sky survey | 3.08 digital<br>1.39 early | |
| 9 | 14 | (17.58)<br>sloan digital sky survey<br>(16.3)<br>commissioning data<br>(12.26)<br>sky survey<br>(11.95)<br>field methane<br>(11.93)<br>high-redshift quasar | (2)<br>field methane dwarf | (42.37)<br>data release<br>(42.37)<br>release | (17)<br>survey commissioning data | (3) five high-redshift quasars discovered in commissioning imaging data of the sloan digital sky survey<br>(3) simulation of stellar objects in sdss color space<br>(3) the discovery of a second field methane brown dwarf from sloan digital sky survey commissioning data<br>(3) the sloan digital sky survey: technical summary<br>(3) topology from the simulated sloan digital sky survey | 4.58 survey<br>4.40 sky<br>3.81 sloan<br>3.81 digital<br>1.87 commissioning | SDSS discoveries: field metane dwarf |
| 3 | 13 | (185.97)<br>generating color<br>(34)<br>existing catalog data | (11)<br>existing catalog data | (249.73)<br>cosmology<br>(208.33)<br>cluster | (121)<br>gravitational lensing | (11) generating colors and k-corrections from existing catalog data | -0.85 data<br>-1.40 digital<br>-1.40 sky<br>-1.40 sloan<br>-1.40 survey | Generating colors |
| 5 | 12 | (152.16)<br>internal velocity<br>(27.82)<br>mass distribution<br>(27.82)<br>cosmogonic model | (9)<br>internal velocity | (33.14)<br>cluster<br>(28.41)<br>clusters | (17)<br>galaxies : clusters : general | (9) internal velocity and mass distributions in simulated clusters of galaxies for a variety of cosmogonic models | 0.00 lensing<br>0.00 time<br>0.00 cosmological<br>0.00 implications<br>0.00 simulations | Cosmogony |
| 18 | 12 | (118.35)<br>first survey<br>(21.64)<br>radio sky | (1)<br>sdss j094857 | (11.99)<br>cosmology:observations<br>(9.96)<br>cosmic virial theorem | (5)<br>bias | (7) the first survey - faint images of the radio sky at 20 centimeters | -0.01 redshift<br>-0.01 lensed<br>-0.01 detect<br>-0.01 gravitationally<br>-0.01 galaxy | Sky surveys |
| 2 | 11 | (28.77)<br>cosmic string<br>(15.46)<br>gravitational lensing signature<br>(15.46)<br>long cosmic string | (1)<br>cosmological constraint | (7.19)<br>reference systems<br>(3.9)<br>methods, data analysis | (3)<br>reference systems | (5) gravitational lensing signature of long cosmic strings<br>(5) observing long cosmic strings through gravitational lensing | 1.73 gravitational<br>1.58 lensing<br>1.54 strings<br>1.54 cosmic<br>1.28 long | Cosmic string |
| 14 | 10 | (37.86)<br>brown dwarf<br>(30.91)<br>optical spectra<br>(27.82)<br>cool brown dwarf | (26)<br>sloan digital sky survey | (22.43)<br>cosmology<br>(18.03)<br>galaxies | (22)<br>large-scale structure of universe | (10) the discovery of a second field methane brown dwarf from sloan digital sky survey commissioning data<br>(10) the near-infrared and optical spectra of methane dwarfs and brown dwarfs | 3.08 sky<br>3.08 survey<br>2.58 sloan<br>2.58 digital<br>1.01 dwarfs | Methane dwarfs and brown dwarfs |

We can make the following observations. The tf*idf selection is often characterized by high-frequency and generic terms, but its power for differentiating clusters is relatively low. The log-likelihood selection is more useful for differentiating clusters, although some terms may be less representative than the tf*idf selection. The LSA-based selection appears to echo the tf*idf selection. Titles appear to be a better source than index terms for the purpose of labeling clusters because index terms tend to be overly broad.

Highlighting co-citation links in consecutive time slices can help analysts to better understand the dynamics of the field of study. For example, as shown in Figure 7, Cluster 8 was highly cited in 2001 by high-redshift quasar papers with a few between-cluster co-citation links connecting the dust emission cluster (#9). In contrast, as shown in Figure 8, co-citation links made in 2005 suggest that the research in 2005 was essentially connecting three previously isolated clusters as opposed to adding within-cluster co-citation links. Cluster 5, background QSO, was cocited with Cluster 10 luminous red galaxy. Cluster 5 was also cocited with Cluster 8 high-redshift quasar.

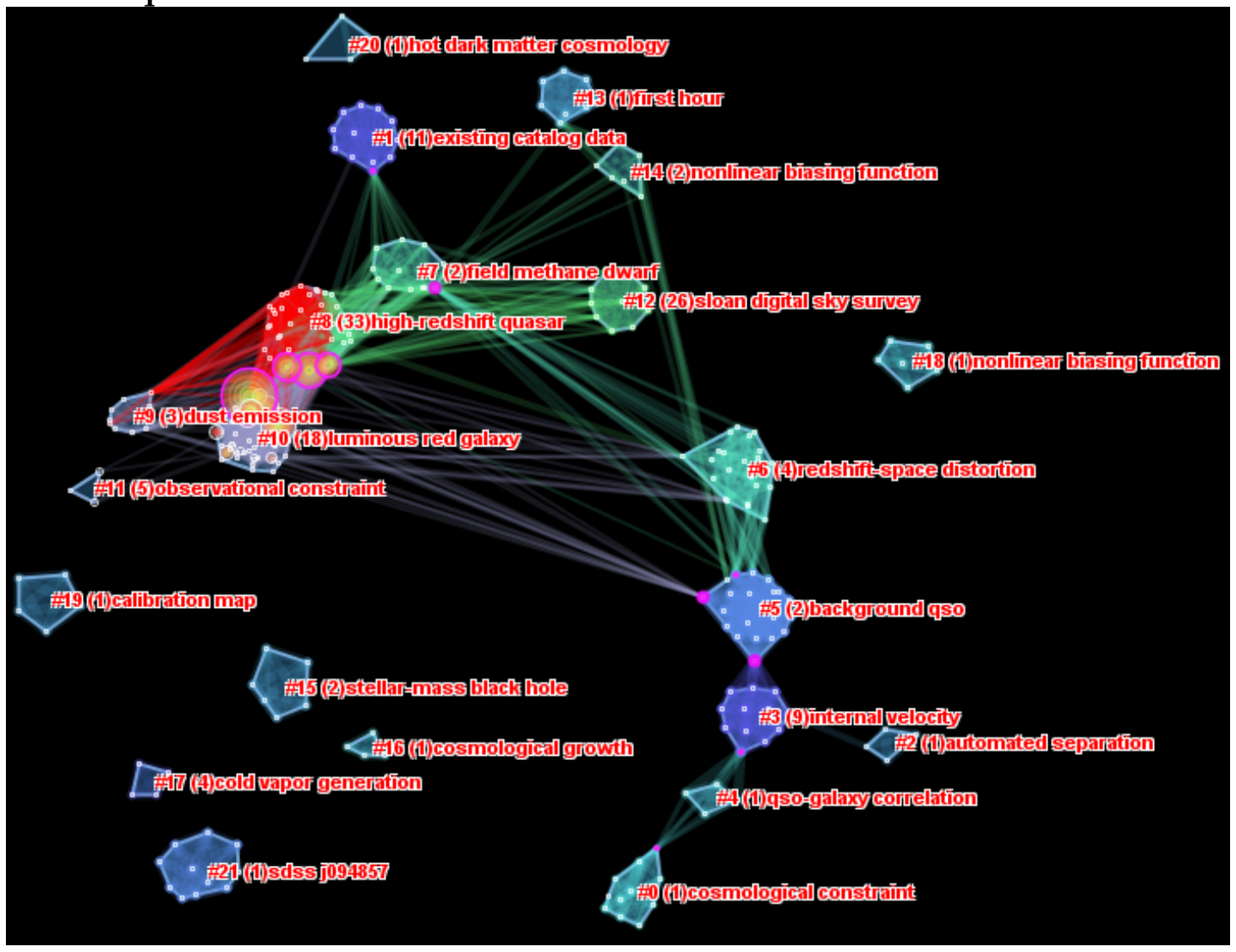

**Figure 7. Co-citation links made in 2001 (in red), primarily in Cluster 8 and linking to Cluster 9.**

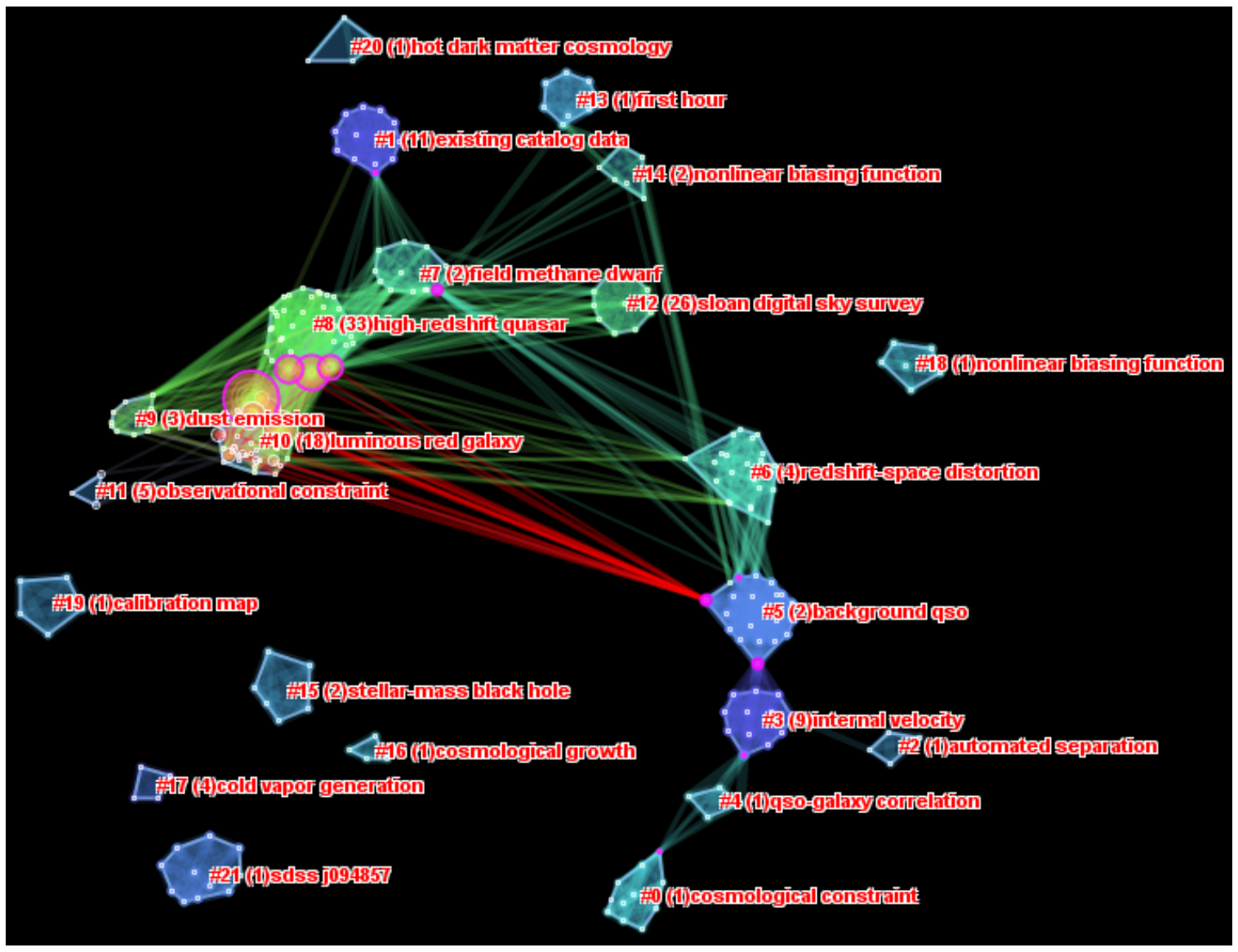

**Figure 8. Co-citation links made in 2005 (in red) between Cluster 10 and Cluster 5.**

### *Clusters in Networks of Co-Occurring Terms*

Various properties of clusters in term networks from arXiv, ADS, and NSF awards are summarized in Table 3. Top 300 most frequent terms were chosen to form the networks for arXiv and ADS. In contrast, top 30 most frequent terms per year were used for the 10-year period of time, which led to 268 unique terms overall. In terms of the length of the time span, arXiv was the shortest, i.e. 15 days, whereas NSF awards were the longest, i.e. 10 years. NSF awards resulted 23 clusters, much smaller than the 52 and 61 clusters of the other two sources. Their modularity measures were about the same between 0.80~0.90. However, NSF awards had the highest the mean silhouette value of 0.8815. ADS had 0.5669 in the middle. arXiv had the lowest 0.4733. These observations suggest that the length of the time span of a dataset might have an influence on the number of clusters. It would be reasonable to conjecture that the longer a time span is, the fewer and larger-sized clusters they would form. The cluster size distribution shown in Figure 9 would be consistent with this conjecture. In addition, the NSF award dataset has consistently higher silhouette values than the other two datasets (See Figure 9 right).

**Table 3. Additional sources on SDSS, including e-prints and funding awards. All clusters are labeled by title terms and ranked by LLR. * The date of search: August 22, 2009.**

| Source | Timespan | Top N terms | Nodes | Links | Clusters | Modularity | Mean Silhouette |
|---|---|---|---|---|---|---|---|
| arXiv | Last 15 days* | 300 | 300 | 1,195 | 52 | 0.8619 | 0.4733 |
| ADS | Jan-Aug* 2009 | 300 | 300 | 525 | 61 | 0.8916 | 0.5669 |
| NSF | 2000-2009 | 30 x 10 slices | 268 | 2,620 | 23 | 0.8275 | 0.8815 |

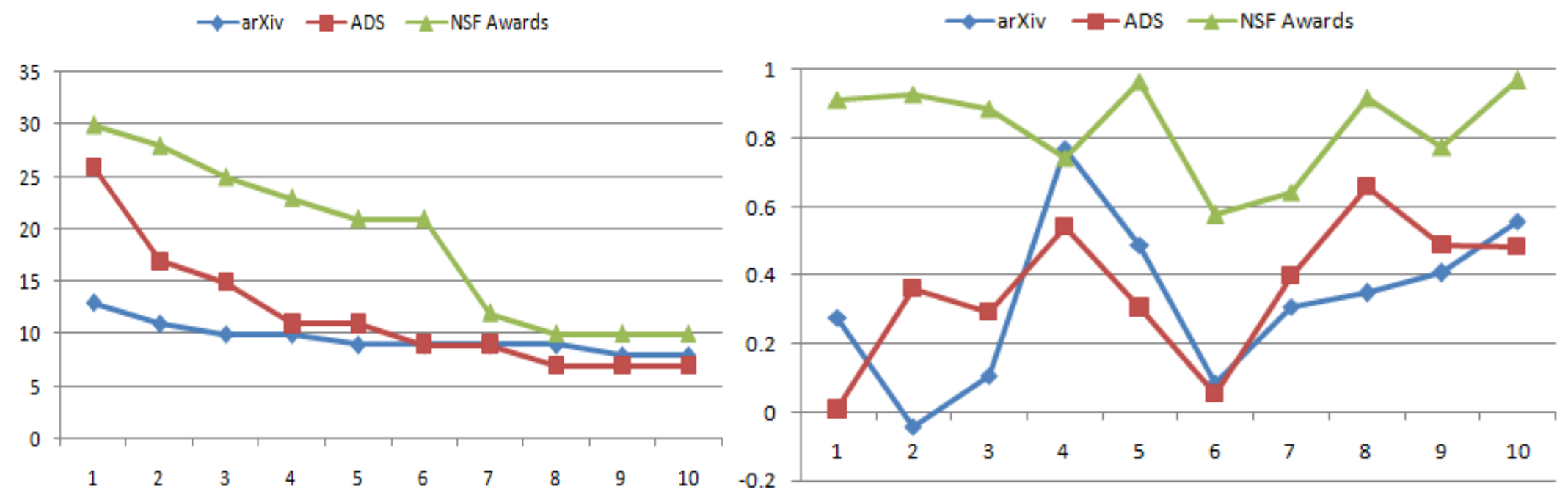

**Figure 9. The cluster size distribution (left) and the mean silhouette distribution (right) of the 10 largest clusters in each of the sources.**

Figures 10-12 show the term networks and clustering structures corresponding to the three different sources. The font size of a cluster label is proportional to the size of the cluster. Figure 10 shows arXiv clusters. Note that they are not necessarily related to SDSS. Instead, they are the latest e-prints posted to the arXiv over the last 15 days. We are interested in learning how our approach will behave on datasets with short as well as long time spans. The arXiv example reveals some interesting properties, especially in contrast to the other two longer-time-span datasets. The ADS network in Figure 11 has the least number of links (525). The *high redshift* cluster (#2) has many terms that are all related to the topic but isolated from each other at this level of selection. The *luminous red galaxy* is the second largest cluster (#47).

The NSF Awards network in Figure 12 is formed by merging 10 networks of each year. The largest cluster is #2, Sloan Digital Sky Survey. The most representative award was made in 2005 by Kron_R, entitled "An extension to the Sloan Digital Sky Survey." According to the color legend, we can tell that the *void galaxy* cluster (#11) and the *galaxy formation* cluster (#16) are the earlier major clusters. In other words, NSF awarded grants in these areas in earlier years. For example, the most representative award in Cluster #11 was made in 2000 by Vogeley_M, a co-author of this article, entitled "voids and void galaxies." The *–phase iii* cluster (#19) was formed in 2008. The most representative award associated to this cluster is by Eisenstein_D, entitled "The Sloan Digital Sky Survey – Phase III." This is apparently the grant to support the SDSS III after the successful SDSS I and II.

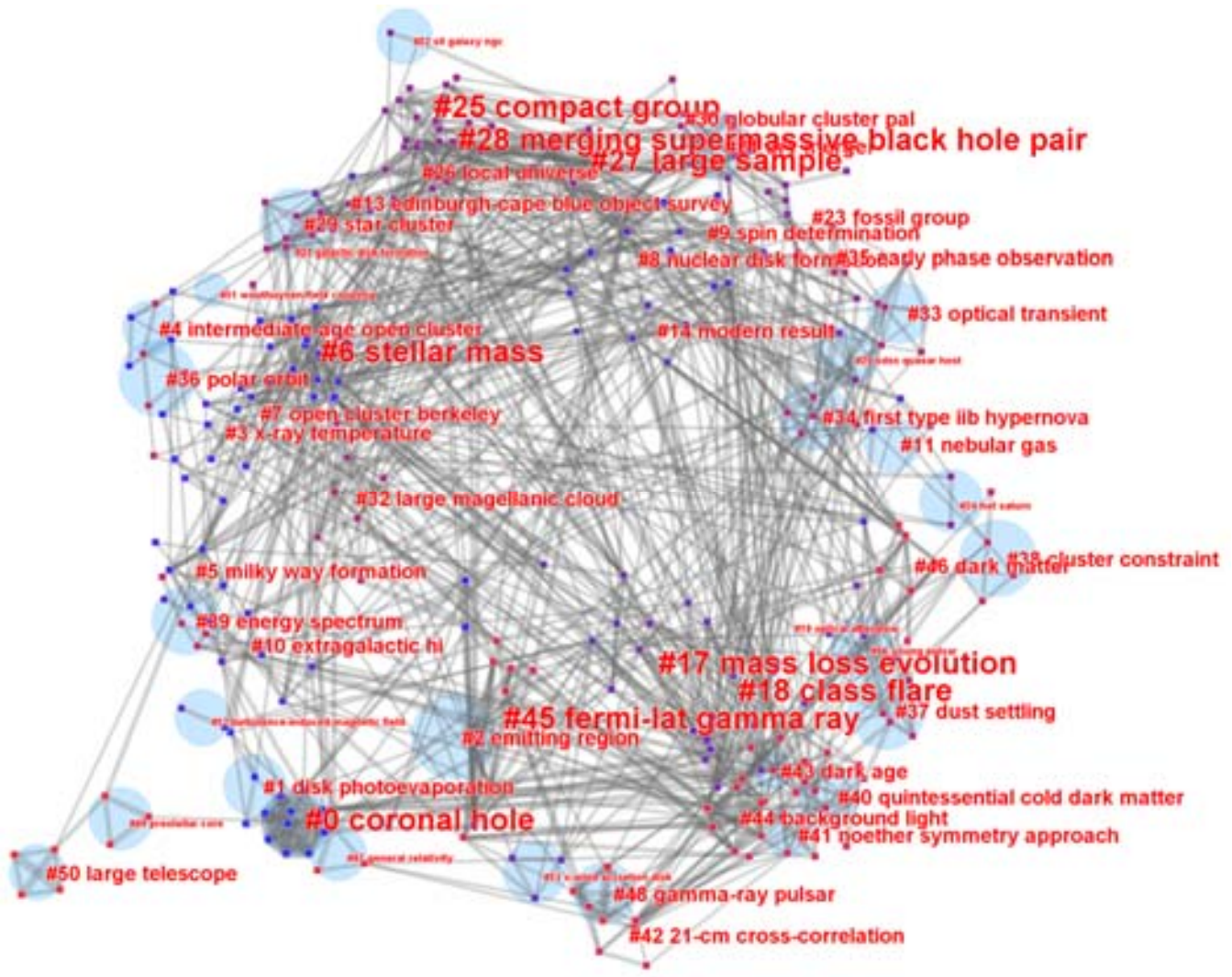

**Figure 10. Clusters of terms in e-prints on astrophysics (astro-ph) submitted to arXiv during the last 15 days of the time of search (August 22, 2009), showing 52 clusters labeled by title terms.**

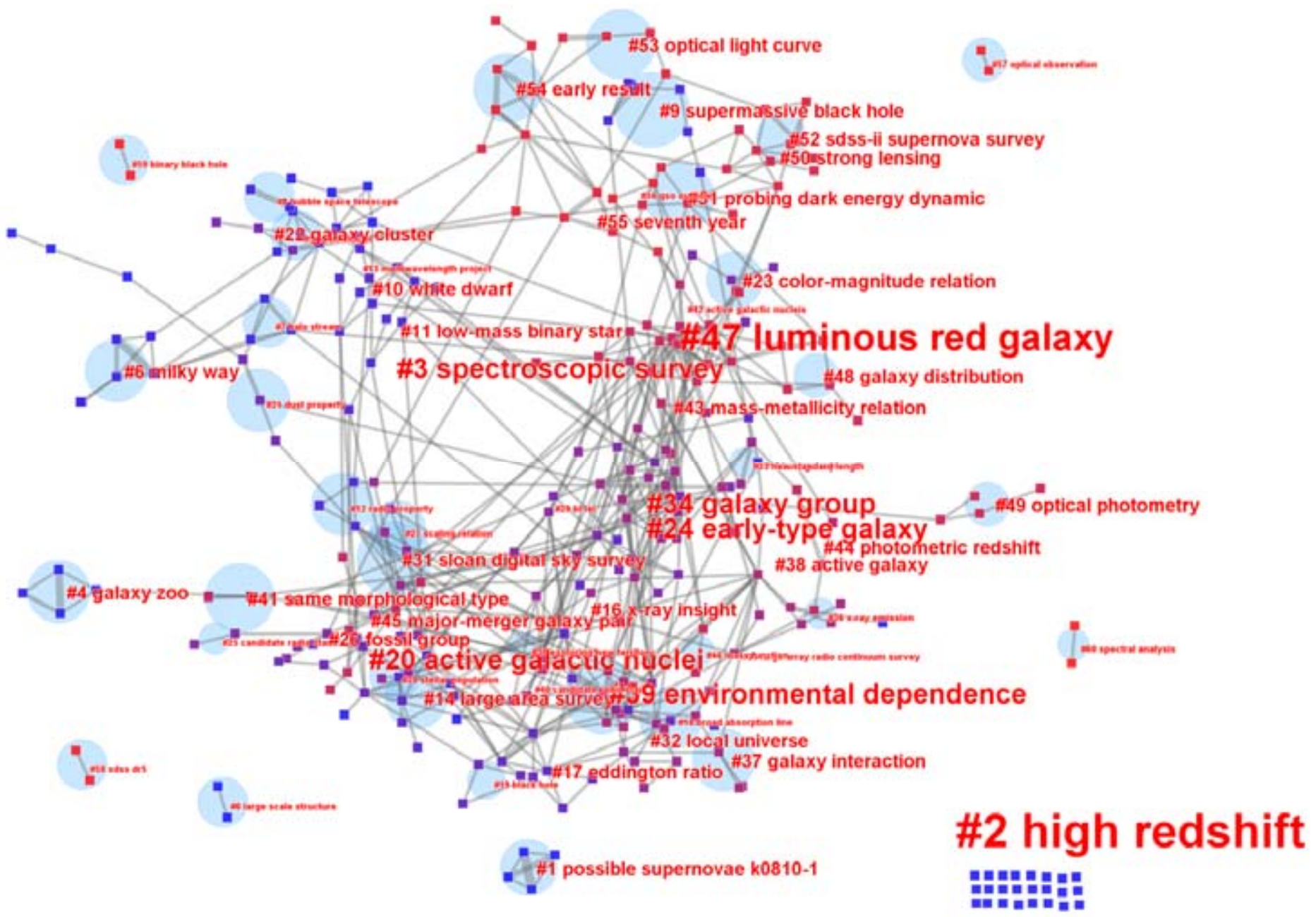

**Figure 11. Clusters of terms in SDSS publications during January and August 22, 2009 in ADS.**

In order to inspect any overlapping clusters across data sources at all, we examined the top 10 largest clusters from each dataset in addition to the largest 10 clusters from the cocitation analysis (See Table 4). The largest two cocitation clusters matched to the largest two ADS term clusters, although in different order. Several cocitation clusters (WoS) are about discoveries made with the SDSS data, e.g. #2, #5, and #10. In arXiv, black hole and dark matter are recurring topics, e.g. #1 and #5, #7 and #10. In ADS, the largest two clusters are the same as the WoS clusters. This is not really a surprise because the ADS dataset is the most similar to the WoS dataset. Clusters in the NSF Awards term network have clearer boundaries as shown in Figure 12 and the high silhouette values. Cluster labels appear to be clear and specific, such as *carbon-enhanced star*, *void galaxy*, *galaxy formation*, and *SDSS weak lensing study*.

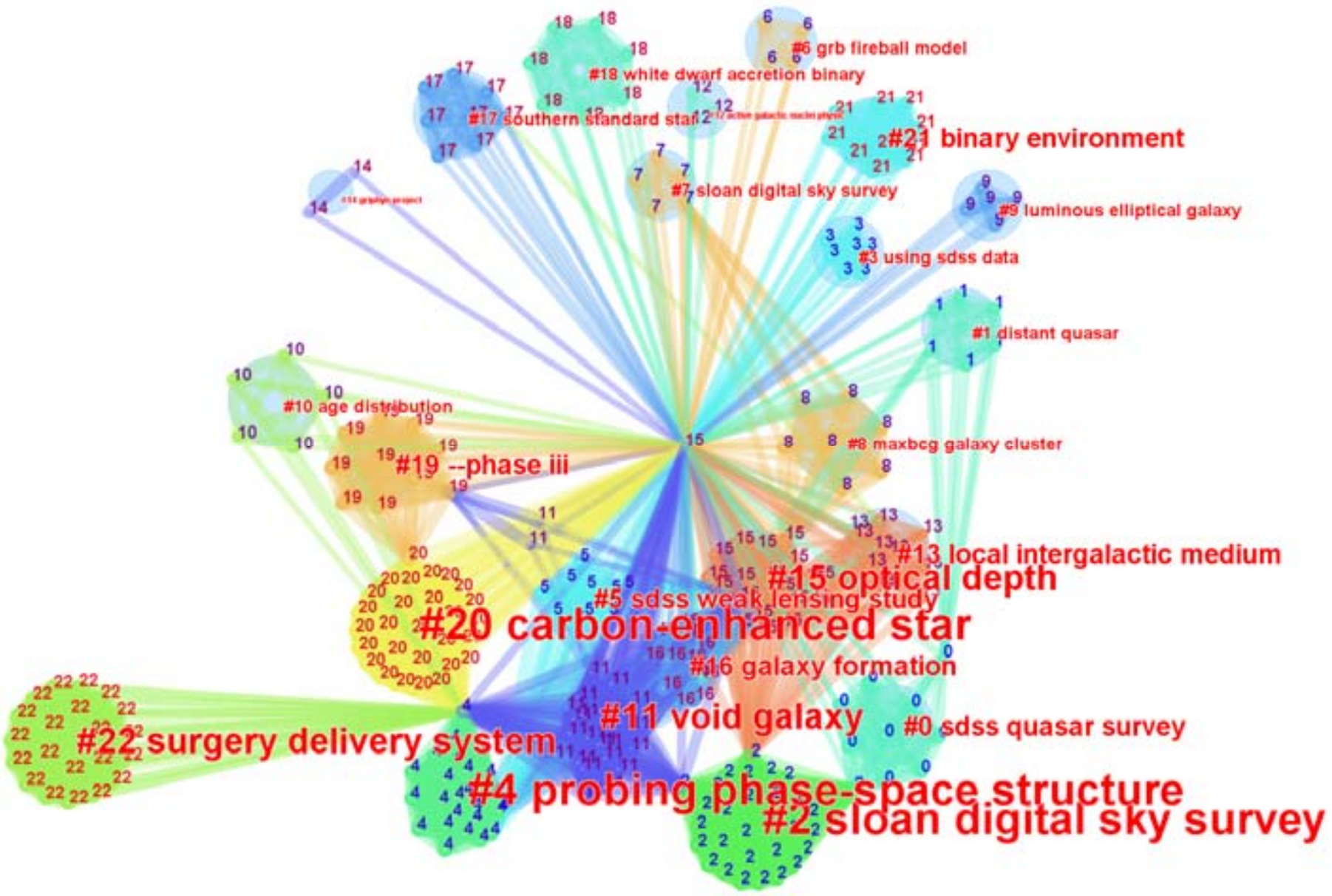

**Figure 12. Clusters of terms in SDSS-related NSF awards.**

**Table 4. Top clusters in four networks on SDSS from heterogeneous data sources.**

| Cluster | WoS [citing context] | arXiv eprints Last 15 days | ADS papers Last 8 months | NSF awards Last 10 years |
|---|---|---|---|---|
| 1 | *luminous red galaxy* [optical properties] | black hole | *high redshift* | SDSS |
| 2 | *high-refshift quasar* [discovery] | x-ray observation | *luminous red galaxy* | carbon-enhanced star |
| 3 | redshift-space distortion [gravitational lensing] | low-metallicity emission-line galaxies | halo model | probing phase-space structure |
| 4 | background QSO | coronal hole | environmental dependence | void galaxy |
| 5 | field methane dwarf [discovery] | merging supermassive black hole pair | color-magnitude relation | surgery delivery system |
| 6 | generating colors | mass loss | spectroscopic survey | optical depth |
| 7 | internal velocity [cosmogonic models] | dark matter signal | active galactic nuclei | galaxy formation |
| 8 | first survey | compact group | scale length | sdss weak lensing study |
| 9 | cosmic string | galaxy cluster | major-merger galaxy pair | local intergalactic medium |
| 10 | brown dwarf and methane brown dwarf | quintessential cold dark matter | local universe | low mass star |

## Discussions and Conclusions

We have introduced a new generic procedure for analyzing the impact of a co-citation network and a term network. The new procedure shifts the focus from cited references to citers to these references and aims to characterize the nature of co-citation clusters in terms of how they are cited instead of inferring how they ought to be cited. Furthermore, the new procedure provides a number of mechanisms to aid the aggregation and interpretation of the nature of a cluster and its relationships with its neighboring clusters. The new methodology is supported by spectral clustering and enhanced network visualization capabilities to differentiate densely connected network components. In order to aid the sense making process further, we integrate multiple channels for the selection of candidate labels for clusters, ranging from saliency-focused term selection to uniqueness-focused selection.

We have also explored the applicability of the new approach to a heterogeneous set of data sources. It is particularly valuable to be able to go beyond typical citation datasets and compare and triangulate patterns found in related but distinct data sources. We have demonstrated that the approach can be consistently applied to time periods as short as days and as long as a decade and even longer. We have also identified some potential properties of time span on the outcomes of clustering and labeling. These properties need to be further investigated in longitudinal studies.

We are addressing some challenging methodological and practical issues. Further studies are needed to evaluate the new method at a deeper level. We have noticed that when our astronomy experts attempted to make sense of bibliographic clusters, they tend to use algorithmically selected terms as a starting point and find concepts at appropriate levels of abstraction. The final concepts they choose may not necessarily present in the original list of candidates. In such synthesizing processes, scientists appear to search for a match in the structure of their domain knowledge. If this is indeed the case, it implies that the primary challenge is to bridge the gap between piecemeal concepts suggested by automatically extracted terms and the more cohesive theoretical organization of the experts. Cluster labels generated by different ranking algorithms may reflect different aspects of a more complex topic. For example, given a phrase searching for dark matter signals in Fermi-lat gamma rays, the tf*idf ranking may choose *dark matter signals*, whereas the log likelihood ratio may choose *Fermi-lat gamma rays*. More sophisticated labeling structure than a single term may capture the nature of a cluster more meaningfully and comprehensively.

Further studies are also necessary to compare with relevant methods such as clustering based on bibliographic coupling (Kessler, 1963; Morris, Yen, Wu, & Asnake, 2003). Comprehensive studies of the interrelationships between different labeling mechanisms are important too. For example, one may examine the positions of various labels of the same cluster in terms of their structural properties in a network of labels. Comparative studies with traditional co-citation network analysis will be valuable to provide the empirical evidence that may establish where the practical strengths and weaknesses of the new approach.

In conclusion, the major contribution of our work is the introduction of a new and integrated procedure for analyzing and interpreting co-citation networks from the perspectives of citers. The new method has the potential to bridge the methodological gap between co-citation analysis and other citer-focused analytic methods. The method is readily applicable to a wider range of sense-making and analytical reasoning tasks with associative networks such as social networks and concept networks by cross-validating structural patterns with direct and focused content information.

## Acknowledgements

This work is supported in part by the National Science Foundation (NSF) under grant number 0612129.

## Notes

CiteSpace is freely available at http://cluster.cis.drexel.edu/~cchen/citespace.